\documentclass[]{aastex631}

\usepackage{subfigure} %

\usepackage{amsmath}

\begin{document}

\title{Local interstellar spectra and solar modulation of cosmic ray proton and Helium}

\author{Cheng-Rui Zhu}
\altaffiliation{zhucr@ahnu.edu.cn}      
\affiliation{Department of Physics, Anhui Normal University, Wuhu 241000, Anhui, China}

\affiliation{Key Laboratory of Dark Matter and Space Astronomy, Purple
Mountain Observatory, Chinese Academy of Sciences, Nanjing 210008, Jiangsu, China}



\begin{abstract}

    The galactic cosmic rays (GCR) suffer from  solar modulation when they propagate through the heliosphere. 
    The transfer of the local interstellar spectrum (LIS) to the top-of-atmosphere spectra (TOA) is influenced by solar wind convection, diffusion on the heliospheric magnetic field (HMF), among other factors. 
    In this work, we derive the LIS of proton (p) and helium (He)  covering energies from a few MeV/n to TeV/n, using a non-parameterization method. The study utilizes monthly AMS-02 data on proton  and helium fluxes and their ratio to examine the evolution of solar modulation from May 2011 to May 2017. To improve the fitting, the force-field approximation is modified by assigning different solar modulation potentials for high ( $\phi_h$ ) and low ($\phi_l$ ) energy ranges. A sigmoid function is employed to describe the transition between these energy ranges. The analysis reveals that the break in proton and helium fluxes occurs at the same rigidity value, with a mean of approximately 6 GV and  this break is more pronounced during the heliospheric magnetic field reversal period. The $\phi_l$ is close to the result of Advanced Composition Explorer (ACE) while the $\phi_h$ is close to the result of neutron monitor (NM) data. Furthermore, the long-term behavior of the p/He ratio is found to naturally arise from the model when considering different Z/A values and the LISs for proton and helium.

\end{abstract}



\section{Introduction} \label{sec:intro}

It is now widely believed that Galactic cosmic rays (GCRs) get accelerated  at cosmic accelerators such as shocks of supernova explosions, and then propagate diffusively in the Galactic random magnetic field \citep{1998ApJ...493..694M}. 
Upon their entry into the heliosphere, interactions with the solar wind and its encapsulated magnetic field induce alterations in both the 
intensity and spectral characteristics of low-energy cosmic rays, 
differentiating them from the local interstellar spectrum (LIS)
\citep{Potgieter2013}. These influences on cosmic rays, termed solar modulation, constrain our comprehension of  cosmic rays. Consequently, examining solar modulation is crucial for exploring the injection and 
propagation of cosmic rays, indirect dark matter detection, and the diffusion theory within the galaxy and heliosphere 
\citep{Yuan:2014pka,Tomassetti:2017hbe,YuanQ_2018}. Furthermore, the fluctuating cosmic ray flux within interplanetary space presents 
substantial challenges for both space missions and atmospheric travelers \citep{Space_Weather}.


The Parker equation is used to   describe the GCRs’ transport processes in the heliosphere, and it can be solved by numerical methods or analytical methods. Under varying levels of approximations, usually assuming spherical symmetry, we can get the force field approximation (FFA) \citep{1967ApJ...149L.115G,1968ApJ...154.1011G} which is widely used to solve the solar modulation as it is simple and enough to {explain most of the observations.}

Now, with the development of instruments, such as PAMELA, AMS-02 and DAMPE \citep {2011Sci...332...69A,2017PhRvL.119y1101A,cite-key} , the observation has entered a high-precision era and these experimental results
are useful to understand the solar modulation effect.  The Voyager 1 flew outside the heliosphere on August 2012 and directly measured the LIS in the range from a few to hundreds MeV/nucleon\citep{2013Sci...341..150S}.  Recently, the AMS-02 measured the time variation of the cosmic ray proton and helium flux between May 2011 and May 2017 in the rigidity range from 1 to 60 GV, at monthly time resolution \citep{PhysRevLett.121.051101}.
{More recently, AMS-02 published the time variation of the daily p and He up to May 2019 \citep{AMS:2021qln,AMS:2022ojy}. In this work, we only use the monthly fluxes from \cite{PhysRevLett.121.051101}. The daily AMS-02 fluxes will be subject of a future study.}

The precise measurement reveals that the Force Field Approximation (FFA) is inadequate to fully account for all the data,  including  the GCR spectra themselves and flux ratios. The p/He flux ratio, has a clear long term trend in time below 3 GV: it remains flat until March 2015, and then it decreases by about 5\% around 2 GV in the next two years. \cite{corti2019time} have explained time dependence of the p/He ratio in cosmic rays using the framework of the force-field approximation.  \cite{PhysRevLett.121.251104,Luo_2019,Corti_2019,Song_2021,2022PhRvL.129w1101Z,2022PhRvD.106f3006W} reproduced the AMS observations using a one-dimensional or a three-dimensional numerical model  respectively to solve the Parker equation.
{Several methods have been proposed to expand the  FFA (\citep{2016ApJ...829....8C,2016PhRvD..93d3016C,2017PhRvD..95h3007Y,2017JGRA..12210964G,Zhu:2020koq,2021ApJ...921..109S,Cholis_2022,PhysRevD.109.083009})  to account for the differences in observed and predicted GCR spectra.}

In this work, we aim to refine the conventional force field approximation by allocating distinct solar modulation potentials to high and low energy cosmic rays, focusing on the time-dependent fluxes of protons and helium. Since the solar modulation is influenced by the Local Interstellar Spectrum (LIS) model, we will employ a non-parametric method to determine LIS by fitting data from Voyager-01 and AMS-02. Subsequently, we will deduce the time-dependent nature of solar modulation using the monthly AMS-02 proton and helium fluxes.

\section{Methodology}\label{Methodology}
\subsection{Solar Modulation}

{GCRs are modulated by} the heliospheric magnetic field carried by
solar winds when they enter the heliosphere, resulting in suppression
of their fluxes. This solar modulation effect depends on particle 
energies, and is particularly obvious at low energies. The Parker's equation \citep{1965P&SS...13....9P} is used to describe the process and the  force-field approximation ( FFA) \citep{1967ApJ...149L.115G,1968ApJ...154.1011G} is a widely used solution.
In the FFA, the TOA flux is related with the LIS flux as
\begin{equation}\label{force_filed}
J^{\rm TOA}(E)=J^{\rm LIS}(E+\Phi)\times\frac{E(E+2m_p)}
{(E+\Phi)(E+\Phi+2m_p)}, 
\end{equation}
where $E$ is the kinetic energy per nucleon, $\Phi=\phi\cdot Z/A$ with 
$\phi$ being the solar modulation potential, $m_p=0.938$ GeV is the 
proton mass, and $J$ is the differential flux of GCRs. The only
parameter in the force-field approximation is the modulation potential $\phi$.

In principle, the force-field approximation assumes a quasi-steady-state of 
the solution of the Parker's equation. 
The observed GCR fluxes display 11-year fluctuations linked to solar activities. Consequently, a time-series of $\phi$ at various epochs is employed to characterize the data. Given that a single parameter is inadequate for accurately fitting the monthly cosmic ray fluxes, we have modified the force-field approximation by assigning distinct potentials to high and low energy cosmic rays. We adopt a sigmoid function to depict the novel solar modulation potential as follows:
\begin{equation}\label{sigmoid}
\phi(R) = \phi_l +\left (\frac{\phi_h-\phi_l}{1+e^{(-R+R_b)}} \right )
\end{equation}
where $\phi_l$ is the solar modulation potential for the low energy, and $\phi_h$ is for the high energy, e is the natural constant, R is the rigidity and $R_b$ is the break rigidity. This model is developed from the model in \cite{2016ApJ...829....8C,2017JGRA..12210964G}.
{\cite{2021ApJ...921..109S} and \cite{PhysRevD.109.083009}  both investigate solar modulation using a modified FFA approach. However, the model proposed by \cite{2021ApJ...921..109S} does not provide a consistent description for protons and helium. On the other hand, the model by \cite{PhysRevD.109.083009} introduces a scaling factor, g, which is associated with the strength of the magnetic field, B, and the solar wind velocity, V. Nonetheless, this constant factor, g, may exert a uniform influence across all energy ranges, which could be physically implausible. This seems more akin to rescaling the LIS, which essentially involves assigning different LIS values to distinct distinct epochs.}

\subsection{LIS of proton and Helium}

Usually power-law or broken power-law functions are employed to fit the 
GCR data \citep{2014A&A...566A.142Y}. If the observational data cover a wide enough energy range, one 
can instead use a non-parametric method by means of spline interpolation 
of GCR fluxes among a few knots \citep{2016A&A...591A..94G,2017AdSpR..60..833G,Zhu:2018jbk}. 
The spline interpolation is a way to obtain an approximate function 
smoothly passing through a series of points using piecewise polynomial 
functions. We use the cubic spline interpolation here, with the 
highest-order of polynomial of three. We work in the $\log(E)-\log(J^{LIS})$ 
space of the energy spectrum, and the units of $E$ is $GeV/n$, the units of $J^{LIS}$ is $m^{-2} s^{-1}sr^{-1} (GeV/n)^{-1}$. The positions of knots of $x=\log(E)$ used here are 
\begin{equation}
\begin{split}
    &\{x_1,x_2,x_3,x_4,x_5,x_6,x_7,x_8,x_9\} \\
    =&\{-3.0,-1.6,-0.9,-0.2,0.5,1.2,1.9,2.6,3.3\}.
\end{split}
\end{equation}
In the following, $y_i$ parameters at the
above fixed $x_i$ knot positions are assumed to be free and are derived 
through fitting to the data.

We fit the normalizations of the $n$ spline knots, together with the
solar modulation potential $\phi$. The $\chi^2$ statistics is defined as
\begin{eqnarray}
\chi^2=\sum_{i=1}^{m}\frac{{\left[J(E_i;\boldsymbol{y},\phi)-
J_i(E_i)\right]}^2}{{\sigma_i}^2},
\end{eqnarray}
where $J(E_i;\boldsymbol{y},\phi)$ is the expected modulated flux, $J_i(E_i)$ and 
$\sigma_i$ are the measured flux and error for the $i$th data bin with  {geometric mean energy $E_i$. }

We use the Markov Chain Monte Carlo (MCMC) algorithm to minimize the
$\chi^2$ function, which works in the Bayesian framework. The posterior
probability of model parameters $\boldsymbol{\theta}$ is given by
\begin{equation}
p(\boldsymbol{\theta}|{\rm data}) \propto {\mathcal L}(\boldsymbol{\theta})
p(\boldsymbol{\theta}),
\end{equation}
where ${\mathcal L}(\boldsymbol{\theta})$ is the likelihood function
of parameters $\boldsymbol{\theta}$ given the observational data, and 
$p(\boldsymbol{\theta})$ is the prior probability of $\boldsymbol{\theta}$.

The MCMC driver is adapted from {\tt CosmoMC} \citep{2002PhRvD..66j3511L,Liu_2012}.
We adopt the Metropolis-Hastings algorithm. The basic procedure of this
algorithm is as follows. We start with a random initial point in the 
parameter space, and jump to a new one following the covariance of these
parameters. The accept probability of this new point is defined as
$\min\left[p(\boldsymbol{\theta}_{\rm new}|{\rm data})/p(\boldsymbol{\theta}_
{\rm old}|{\rm data}),1\right]$. If the new point is accepted, then repeat
this procedure from this new one. Otherwise go back to the old point.
For more details about the MCMC one can refer to \citep{MCMC}.


{Using the modified FFA will cause a greater degeneracy between the LIS and solar modulation parameter with respect to the FFA. Note that the differences in the modulated spectrum induced by $\phi_l$ and $\phi_h$ not being equal are absorbed in the LIS spectrum. For this reason, using the FFA to determine the LIS might slightly bias the LIS values in the energy range where the transition between $\phi_l$ and $\phi_h$ occurs.}
 The GCR data from AMS-02 \citep{AGUILAR20211}, and Voyager-1 \citep{2016ApJ...831...18C} are adopted to fit the LIS. { In 2012, Voyager-1 traversed the heliosphere and subsequently detected the LIS of cosmic rays for the first time.} So in this work, we assume the  solar modulation potential of Voyager data be 0.   AMS-02 provides the most accurate  cosmic rays data in wide energy range and the $\phi^{AMS}$ is the free parameter to be fitted. {With these two datasets we can precisely constrain the LIS.} 
 
\begin{figure}[htbp]	
	\subfigure[] 
	{
		\begin{minipage}{9cm}
			\centering         
			\includegraphics[scale=0.6]{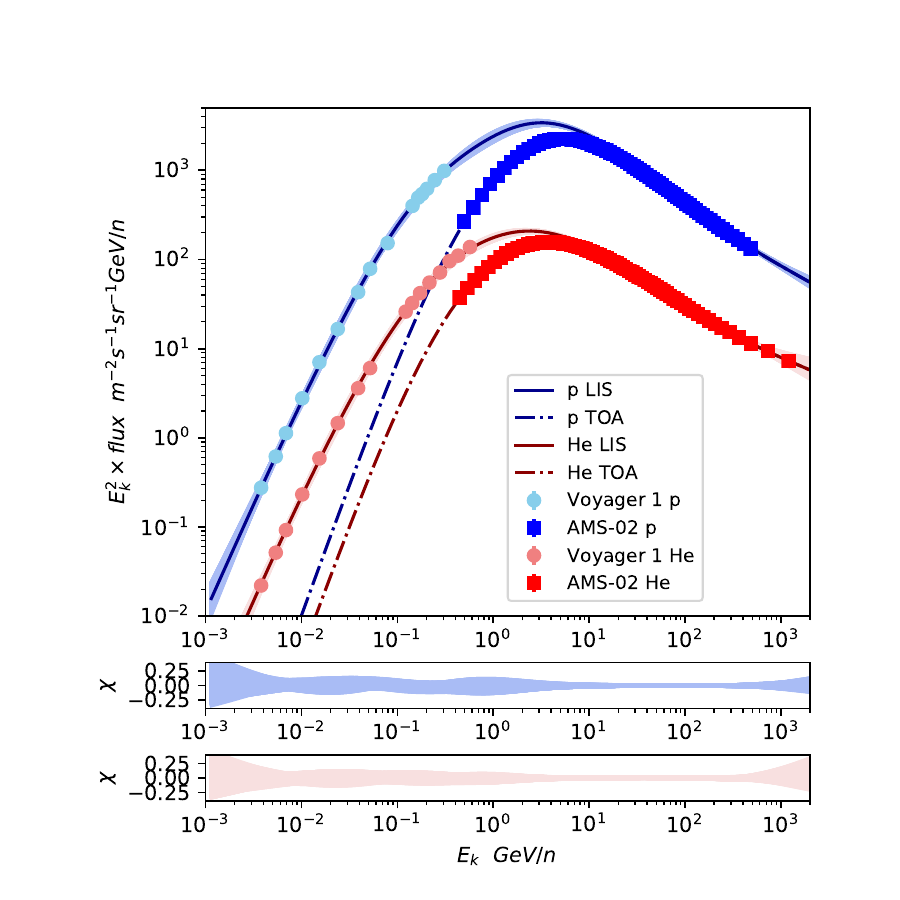}  
		\end{minipage}
	}
	\subfigure[] 
	{
		\begin{minipage}{6cm}
			\centering      
			\includegraphics[scale=0.5]{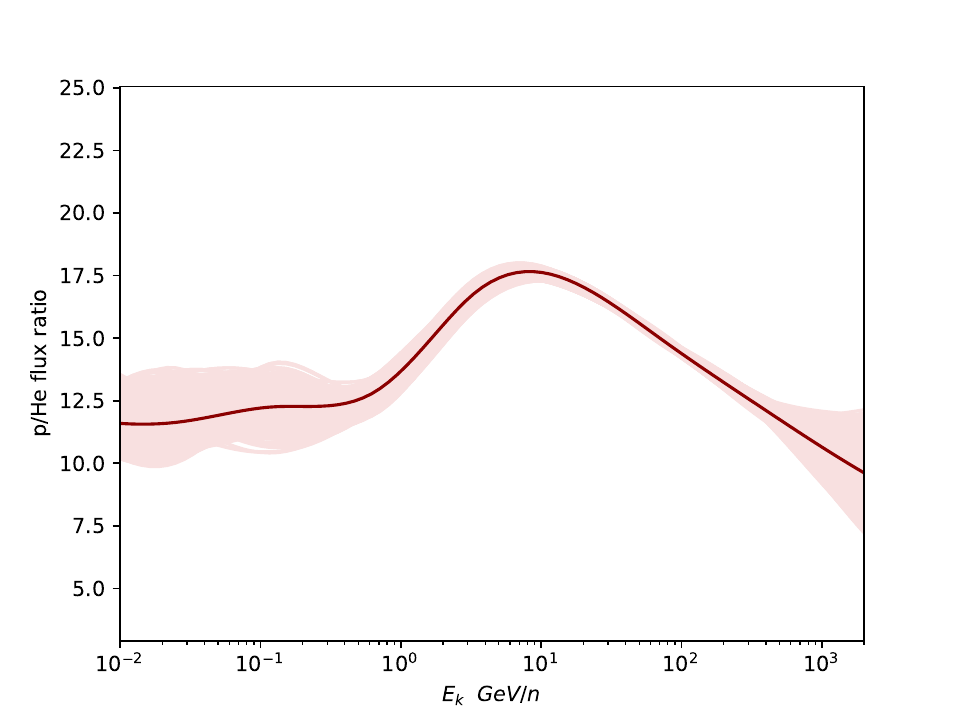}  
		\end{minipage}
	}
	\caption{ {(a) Top panel: Best-fit LIS fluxes (lines) and TOA fluxes (dashed lines ) of p and He, multiplied by   $E_k^{2}$, compared with the measurements (colorful points)     of Voyager 1 \citep{2016ApJ...831...18C}, and AMS-02 \citep{AGUILAR20211}. Bands show the results with the 95\% ranges of the spectral parameters. The middle (bottom) panel shows the the 95\% credible intervals centered around $J_{best}$ for proton (Helium).  (b)  LIS ratio of  proton-to-Helium, the band stands for the 95\% CI.}}  
	\label{fig:LIS}
\end{figure}

  With the help of MCMC, the best fit LIS of proton and Helium are shown in FIG. \ref{fig:LIS}, with solar modulation potential $\phi^{AMS} = 0.477 \pm 0.063 GV$ and $\chi^2/d.o.f = 27.0/121$. Notably, here we fit proton and Helium together with the same $\phi$, if we fit  'p-only’ and ‘He-only’, we will get $\phi^{AMS}_p = 0.445 \pm 0.063$ and $\phi^{AMS}_{He} = 0.494 \pm 0.074 GV$,  which are compatible within their  1 $\sigma$ credible intervals (CI). {In this study, we assume Z=1 and A=1 for proton, Z=2 and A=4 for Helium.}

\section{results}

\begin{figure}[!ht]
    \centering
    \includegraphics[scale=0.8]{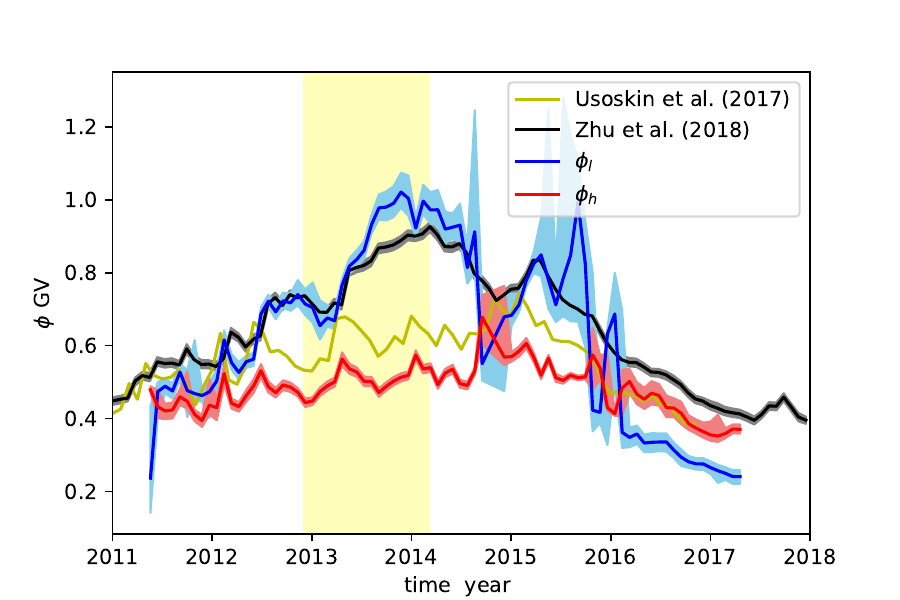}
    \caption{Time series of $\phi_l$ (red line)  and $\phi_h$ (blue line) via fitting to the AMS-02 p, He and p/He from 2011 to 2017, with 1 $\sigma$ bands. Previous results inferred from the
    neutron monitor (NM) data \citep{2011JGRA..116.2104U,2017JGRA..122.3875U} (yellow line ) and GCR data \citep{Zhu:2018jbk}  (dark line) are also shown for comparison. The yellow band stands for the   heliospheric magnetic field reversal period within which the polarity is uncertain\citep{2015ApJ...798..114S}.}
    \label{fig:phi}
\end{figure}

\begin{figure}[!ht]
    \centering
    \includegraphics[scale=0.8]{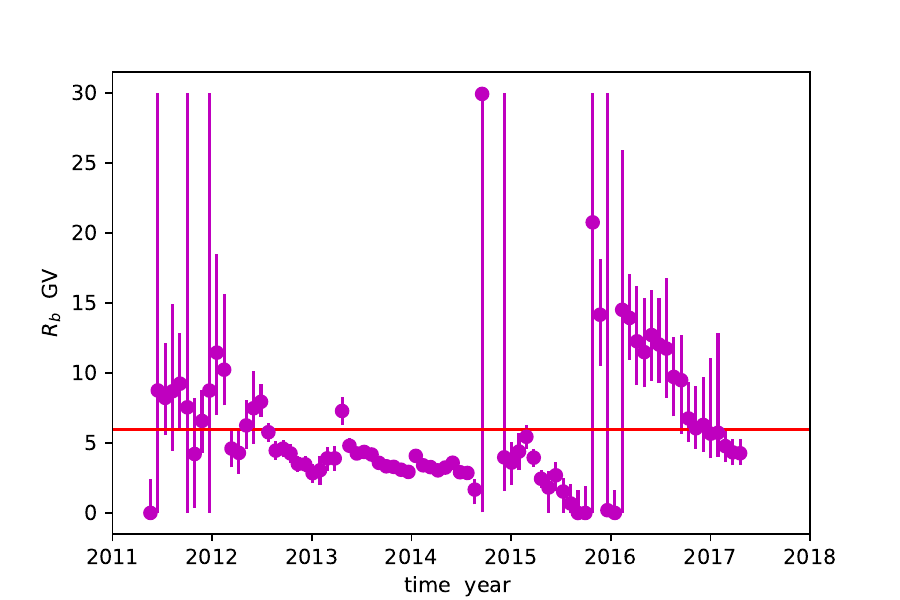}
    \caption{ The rigidity break for different period. The red line stands for the mean value, which is about 6 GV. }
    \label{fig:rb}
\end{figure}

{Upon determining the LIS fluxes of cosmic rays, we can subsequently derive the temporal evolution of the solar modulation by simultaneously utilizing the extensive measurements from AMS-02 with MCMC. This includes data on monthly protons, helium, and the proton-to-helium ratio.}
 {To properly take into account
the uncertainties of the LIS, we adopt a Bayesian approach with the
posterior probability of $\phi_{l}$, $\phi_{h}$ and $R_b$ being given by
\begin{equation}
p(\phi_{l} ,\phi_{h}, R_b|{\rm data})\propto\int {\mathcal L}(\phi_{l} ,\phi_{h}, R_b,\boldsymbol{y} )\,p(\boldsymbol{y})\,{\rm d}\boldsymbol{y},
\end{equation}
where ${\mathcal L}$ is the likelihood of model parameters 
($\phi_{l} ,\phi_{h}, R_b,\boldsymbol{y} $), $p(\boldsymbol{y})$ is the prior 
probability distribution of $\boldsymbol{y}$ which is obtained in the 
fit in Sec.~2.2.}
The result is depicted in Figure \ref{fig:phi}, where the blue line represents $\phi_l$  for low energy, and the red line represents $\phi_h$ for high energy. Considering the uncertainties associated with the LIS, we have depicted the 1 $\sigma$ confidence intervals (CI) with their respective bands. 
The value of $\phi_l$ is consistent with the results from \cite{Zhu:2018jbk}, which were derived from tens MeV/n of Boron, Carbon, and Oxygen as measured by the Advanced Composition Explorer (ACE). Meanwhile, the value of $\phi_h$ has a similar time dependence with the findings of \cite{2011JGRA..116.2104U}, obtained from neutron monitor (NM) data.
The rigidity break, $R_b$ is illustrated in Figure \ref{fig:rb}, with a mean value of 5.96 GV. 
{The values of $R_b$ are consistent with the values of the rigidity break in the diffusion coefficient from \cite{Song_2021} This suggests that the rigidity at which the transition between $\phi_l$ and $\phi_h$ occurs is related to the rigidity break in the diffusion coefficient.}
The $\phi_l$ is higher than $\phi_h$ before 2016, and $\phi_l$ is lower than $\phi_h$ after that. {This is also similar to the result of \cite{Song_2021}.} \cite{Song_2021} shows that, during 2011 -2016, {the high-rigidity power index of the diffusion coefficient $b$ bigger than the index of diffusion coefficients for low-rigidity $c$.}  While after 2016, the situation was completely opposite with $c > b$. 
{This seems to suggest that $\phi_l $and $\phi_h$ are related to the low- and high-rigidity power indices of the diffusion coefficient, respectively.}

\begin{figure}[!ht]
    \centering
    \includegraphics[scale=0.8]{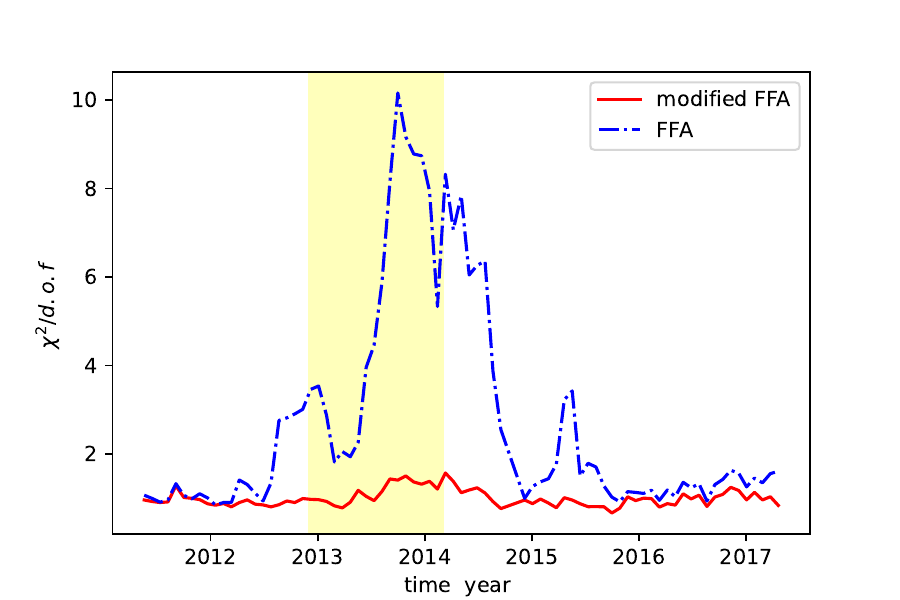}
    \caption{ The  $\chi^2/d.o.f$ from our calculations. The cyan line represents the results obtained using the modified FFA, while the blue line corresponds to the outcomes of the  FFA.The yellow band stands for the  heliospheric magnetic field reversal period within which the polarity is uncertain.}
    \label{fig:chi2}
\end{figure}

The $\chi^2/d.o.f$ is displayed in Figure \ref{fig:chi2}.  The modified FFA yields  $\chi^2/d.o.f$  values ranging between 0.669 and 1.573, with a mean value of 1.006. Compared to the traditional Force Field Approximation (FFA), our model significantly reduces the $\chi^2/d.o.f$, particularly during periods of heliospheric magnetic field reversal when the polarity is uncertain and the  FFA struggles to provide a good fit. However, when the polarity is well-defined, the  FFA can adequately fit the data before 2012/4 and after 2016/10.
{The  $\chi^2$ values from our fitting are primarily influenced by the data within the rigidity range of 4 to 10 GV which is also the range of values for $R_b$. This suggests that the deviations between the model and the data are most significant in this energy range, which may indicate areas for further model refinement or potential complexities in the underlying physical processes affecting cosmic rays within this specific energy interval, such as the rigidity-dependence of the $\phi(R)$ or the diffusion coefficient of GCRs. }

\begin{figure}[!ht]
    \centering
    \includegraphics[scale=0.8]{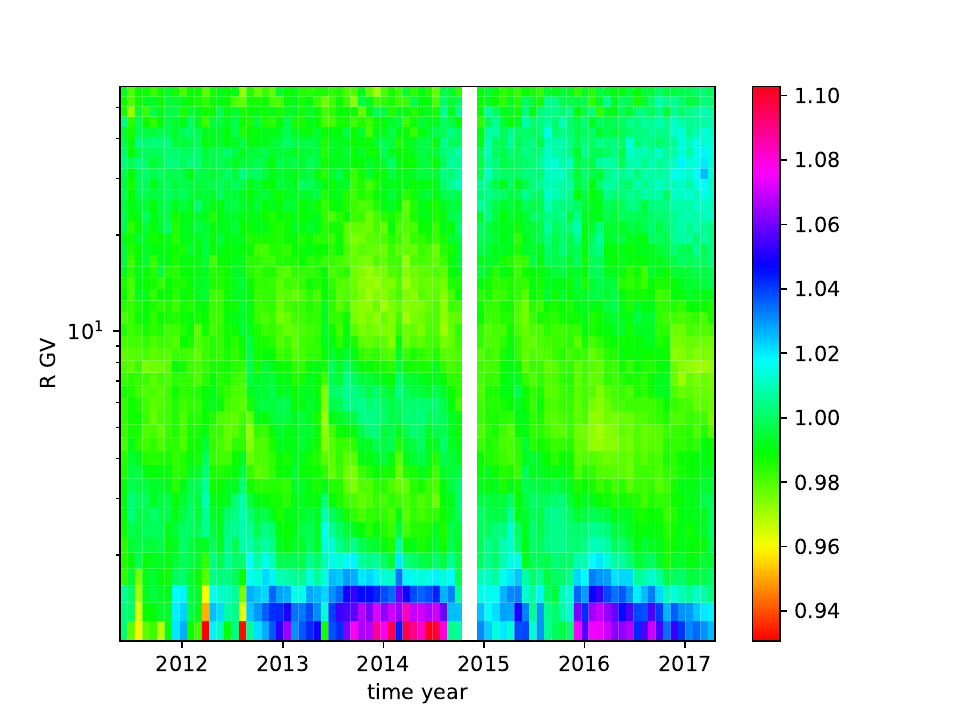}
    \caption{ The ratio of model prediction to data (model/data) for p fluxes of AMS-02 from  may 2011 to may 2017 using modified FFA.}
    \label{fig:p}
\end{figure}

\begin{figure}[!ht]
    \centering
    \includegraphics[scale=0.8]{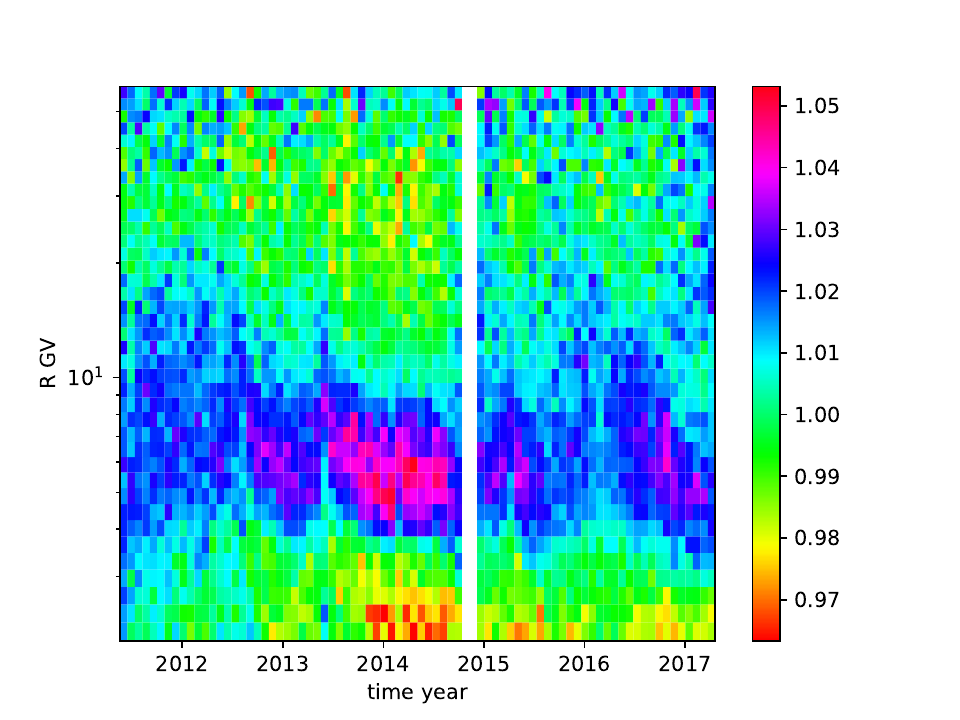}
    \caption{ Same as \ref{fig:p} but for He. }
    \label{fig:He}
\end{figure}

\begin{figure}[!ht]
    \centering
    \includegraphics[scale=0.8]{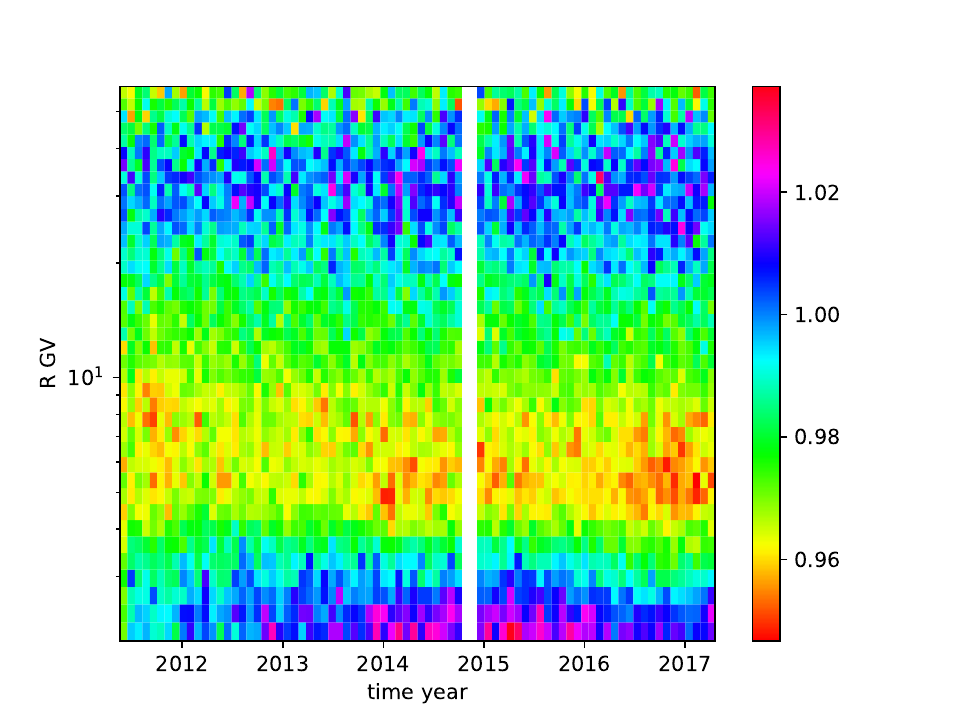}
    \caption{ Same as \ref{fig:p} but for p/He. }
    \label{fig:pHe}
\end{figure}

In Figures \ref{fig:p}, \ref{fig:He}, and \ref{fig:pHe}, we show the ratio of the computed intensities to  AMS-02 measured values (model/data) from May 2011 to May 2017 using the modified FFA.
{These figures show the predicted fluxes for protons, helium, and the ratio of protons to helium, respectively, compared against the actual measurements from AMS-02.}
{Below approximately 2 GV, the model forecasts a higher abundance of protons and a lower abundance of helium. This discrepancy could potentially be attributed to biases in the LIS. Significantly, this contrasts with the findings observed within the 4 to 10 GV rigidity span.}
The modulated fluxes of cosmic rays are indeed very sensitive to the spectral shape and values of their respective Local Interstellar Spectrum (LIS). As noted by \cite{Ngobeni:2020quz}, even small variations in the LIS can significantly impact the modeled fluxes. For instance, within the rigidity range of 4-10 GV, the model predicts Helium fluxes about 5\% higher than the observed data. { If we adjust the LIS for Helium in this range (4-10 GV) by just 2\% lower—a change well within the 95\% error margins of the LIS— it can lead to a substantial reduction in the $\chi^2$ value. } With such an adjustment, the  $\chi^2/d.o.f$ could be reduced to a range between 0.363 and 0.998, with a mean value of 0.580. This demonstrates the significant influence that precise LIS values have on the accuracy of cosmic ray modulation models and highlights the importance of refining our understanding and measurement of the LIS for improved model predictions. {The observed effects could also potentially be attributed to the isotopic composition of Helium, which has not been taken into account in this study. }

\begin{figure}[htbp]	
	\subfigure[] 
	{
		\begin{minipage}{9cm}
			\centering         
			\includegraphics[scale=0.5]{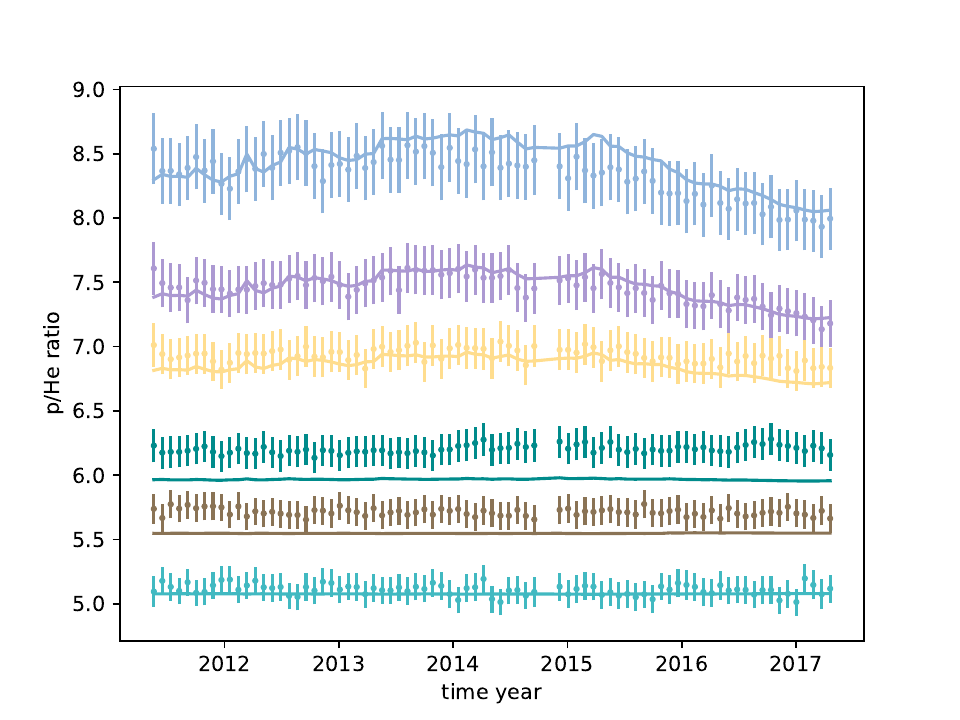}  
		\end{minipage}
	}
	\subfigure[] 
	{
		\begin{minipage}{9cm}
			\centering      
			\includegraphics[scale=0.5]{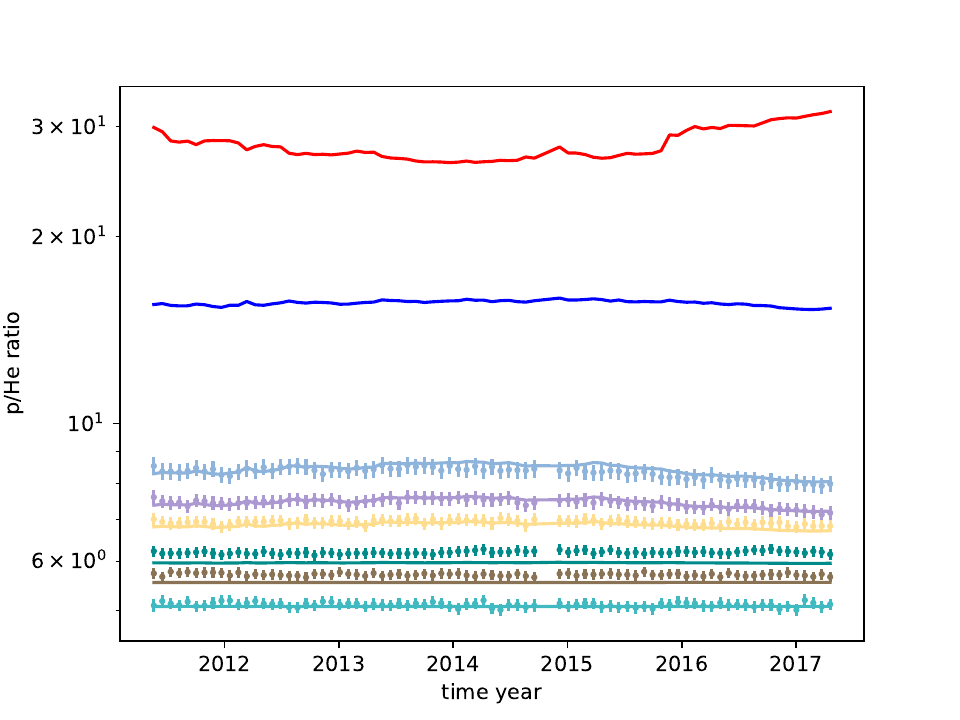}  
		\end{minipage}
	}
	\caption{(a) The model  best-fit time profiles of the p/He ratio evaluated at rigidities R = 2.0 GV, 2.5 GV, 3.1 GV,  5.6 GV, 9.6 GV and 21.9 GV (from top to bottom) compare with the data.  (b) Extending to rigidities R = 0.5 GV (red) and 1.0 GV (blue).} %
	\label{fig:ratio}
\end{figure}

In Figure \ref{fig:ratio}, we present the time profile of the proton-to-helium ratio (p/He) for several rigidity values. This comparison focuses on the best-fit calculations obtained using the modified FFA  against actual data. Using consistent solar modulation parameters, we achieve an accurate fit for p/He.
For the  low rigidity,  the p/He increase from 2011 to 2014 with increasing of solar activity, and then decrease to low values in 2014–2017. The long-term behavior of the p/He ratio is caused by two reasons: {the different Z/A value and LIS for p and He \citep{2022cosp...44.1291S,Song_2023}. }
According to the Eq. \ref{force_filed}, we have the ratio of p and He as:
\begin{equation}\label{eq:ratio}
\begin{aligned}
\frac{J_p^{TOA}(R^{TOA})}{J_{He}^{TOA}(R^{TOA})} =& \frac{J_p^{TOA}(E_p^{TOA}) (\frac{Z\beta}{A})_p}{J_{He}^{TOA}(E_{He}^{TOA}) (\frac{Z\beta}{A})_{He}}\\
 =& \frac{(\frac{Z\beta}{A})_p}{(\frac{Z\beta}{A})_{He}}  \frac{J_p^{LIS}(E_p^{LIS})}{J_{He}^{LIS}(E_{He}^{LIS})} \left ( \frac{R_{He}^{LIS}}{R_p^{LIS}} \right ) ^2.
 \end{aligned}
\end{equation}
Here, $E^{TOA} = E^{LIS } - \frac{Z}{A}\phi(R)$, and $E = \sqrt{R^2(\frac{Ze}{A})^2+m_0^2}-m_0$. 
{For the third term of  Eq. 7, we have
\begin{equation}
\frac{R_{He}^{LIS}}{R_p^{{LIS}}} = 
\frac{\left (\frac{A}{Ze}\right )_{He}}{\left (\frac{A}{Ze}\right )_{p}} 
\frac{\sqrt{ (\sqrt {R^2\left(\frac{Ze}{A}  \right)_{He}+m_0}-m_0+ (\frac{Ze}{A})_{He}\phi(R)) (\sqrt {R^2\left(\frac{Ze}{A}  \right)_{He}+m_0}-m_0+ (\frac{Ze}{A})_{He}\phi(R)+2m_0) } }
{\sqrt{ (\sqrt {R^2\left(\frac{Ze}{A}  \right)_{p}+m_0}-m_0+ (\frac{Ze}{A})_{p}\phi(R)) (\sqrt {R^2\left(\frac{Ze}{A}  \right)_{p}+m_0}-m_0+ (\frac{Ze}{A})_{p}\phi(R)+2m_0) }}.
\end{equation}
For the modified FFA, both p and He share the same solar modulation potential, $\phi(R)$. In contrast, it is challenging to fit p and He simultaneously using the same solar modulation potential within the framework of the FFA. {Because p and He have different Z/A, so the value of $R_{He}^{LIS}/R_{p}^{LIS}$ will change with the $\phi(R)$ time series.}
For the second term, situation is more complicated. We have 
\begin{equation}
 \frac{J_p^{LIS}(E_p^{LIS})}{J_{He}^{LIS}(E_{He}^{LIS})}
=
\frac{J_p^{LIS} \textbf{(}\sqrt{R^2(\frac{Ze}{A})_p+m^2_0}-m_0+(\frac{Z}{A})_p\phi(R)\textbf{)} }
{J_{He}^{LIS}\textbf{(}\sqrt{R^2(\frac{Ze}{A})_{He}+m^2_0}-m_0+(\frac{Z}{A})_{He}\phi(R)\textbf{)}}.
\end{equation}
On the one hand, due to the differing Z/A of p and He, the value within the bold bracket will fluctuate in response to changes in the solar modulation potential, $\phi(R) $, over time. This fluctuation influences the long-term behavior of the ratio between proton and helium fluxes, $J_p/J_{He}$. On the other hand, the spectral shape of the LIS for protons and helium differ, as demonstrated in the Fig.\ref{fig:LIS}. At lower energies (below 0.5 GeV/n), the spectra are similar. { However, at higher energies (but below 10 GeV/n), the proton spectrum is harder than that of helium. Consequently, this leads to an inverse variation in the p/He ratio above and below 1 GV, as depicted in the Fig. \ref{fig:ratio}. }
Therefore, since p and Helium have different Z/A and LIS, the long-term behavior of the p/He ratio arises naturally from the second and third term of Eq. \ref{eq:ratio} with the same $\phi$ series. }


\section{conclusion and discussion}

We study the solar modulation  of p and He with the newly AMS-02 data based on a modified FFA. Using the data of Voyager 1 and AMS-02, we derived the LIS of p and He  with a non-parametric method by means of spline interpolation. Then we fit the solar modulation of AMS-02 time-series of p and He fluxes. With the sigmoid function to replace the constant solar modulation potential in the  FFA, we can fit the data very well.
The sigmoid function have three parameters, $\phi_l$ for the low energy, $\phi_h$ for high energy, and $R_b$ for the break rigidity.  {$\phi_l$ and $\phi_h$ are consistent with $\phi$ from FFA derived using low-energy data (ACE) and high-energy data (NMs), respectively.} The $\phi_l$ is higher than $\phi_h$ before 2016, and $\phi_l$ is lower than $\phi_h$ after that. The break $R_b$ appears at about 6 GV. The break are probably caused by the break of the diffusion coefficient.

{There is a slightly bias  between our model prediction and the data for the 4-10 GV.  The bias of the LIS may be the main reason. If we reduce the LIS of Helium in the 4-10 GV range by 2\% , which is is within the 95\% error range, we can obtain very good fitting results by reduce about half of the $\chi^2$. Notably, the drift effect constitutes another significant factor influencing solar modulation; however, it is not taken into account in the present analysis. This may be one of the origins of the $\chi^2$ observed here.  The isotopic composition of Helium, which has not been considered in this study, could also be one of the factors contributing to the  $\chi^2$ values.}


The modified FFA can interpret the long-term behavior of the p/He ratio recently observed by AMS-02. As the proton and Helium have different Z/A and LIS, the  long-term behavior of the p/He ratio arises naturally from the second and third term of Eq. \ref{eq:ratio} with the same $\phi$ series.

Our model can give a very well fitting to the monthly data of AMS-02 p, He and p/He.  It will be useful to study the origin and  propagation of cosmic rays in the galaxy, and help us understanding the physical of cosmic rays.  We do not consider the charge-sign dependence and drift effect of solar modulation here, we may do that in the future.

\begin{acknowledgments}
Thanks for Qiang  Yuan for very helpful discussions. This work is supported by the National Natural Science Foundation of China  (No. 12203103). C.R.Z is also support by the Doctoral research start-up funding of Anhui Normal University. We acknowledge the use of  data from the \href{https://tools.ssdc.asi.it/CosmicRays/}{Cosmic-Ray Database }\citep{DiFelice:2017Hm}.
\end{acknowledgments}

\bibliography{sample631}{}
\bibliographystyle{aasjournal}



\end{document}